\documentclass[aps,prl,twocolumn,groupedaddress,showpacs]{revtex4}

\usepackage{graphicx}
\usepackage{dcolumn}
\usepackage{bm}

\begin{document}

\title{Odd-Parity Triplet Pair Induced by Hund's Rule Coupling}

\author{Takashi Hotta$^1$ and Kazuo Ueda$^2$}

\affiliation{
$^1$Advanced Science Research Center,
Japan Atomic Energy Research Institute,
Tokai, Ibaraki 319-1195, Japan \\ 
$^2$Institute for Solid State Physics,
University of Tokyo, Kashiwa, Chiba 277-8581, Japan}

\date{\today}

\begin{abstract}
We discuss microscopic aspects of odd-parity triplet pair
in orbital degenerate systems.
From the concept of off-diagonal long-range order, a pair state
is unambiguously defined as the eigenstate with
the maximum eigenvalue of pair correlation function.
Performing this scheme by a numerical technique,
we clarify that the odd-parity triplet pair occurs
as an out-of-phase combination of local triplets
induced by Hund's rule coupling for the lattice including
two sites in the unit cell.
\end{abstract}


\maketitle


Recently triplet superconductivity has been found in several
compounds with strong electron correlation.
A pioneering material should be UPt$_3$ \cite{UPt3},
which has been considered to have spin-triplet pairing
from experimental evidence for multi superconducting phases
similar to the phase diagram of $^3$He.
The odd-parity triplet pair has been eventually confirmed by
the experimental fact that the Knight shift does not change
through the superconducting transition temperature $T_{\rm c}$ \cite{Tou}.
In other $f$-electron compounds, triplet superconductivity
has been also suggested.
For instance, coexistence of superconductivity and ferromagnetism
has been discovered in UGe$_2$ \cite{Saxena} and URhGe \cite{Aoki}.
It is naively believed that spin triplet pair appears
in the ferromagnetic (FM) phase.
Recently UNi$_2$Al$_3$ has been also considered as a triplet superconductor
from NMR measurements \cite{Ishida}.

Among transition metal oxides, Sr$_2$RuO$_4$ has attracted much attention
since its discovery in 1994 \cite{Maeno}.
It is confirmed that this material has spin triplet pair
from NMR measurements \cite{Ishida2},
although it is isostructural to La$_2$CuO$_4$ which is the mother
compound of high-$T_{\rm c}$ cuprates
with singlet $d$-wave superconductivity.
In the $d$-electron system such as ZrZn$_2$, it has been also
reported that superconductivity appears in the FM phase \cite{ZrZn2}
and spin-triplet pairing
is expected to occur in this material.

Regarding the mechanism of triplet superconductivity,
FM spin fluctuation was considered to mediate
the triplet pair also in strongly correlated materials
by analogy with superfluid $^3$He.
However, paramagnons are not always dominant
in the spin fluctuation spectrum of those materials.
In fact, in Sr$_2$RuO$_4$, significant enhancement
of the incommensurate antiferromagnetic spin fluctuation has been
observed in neutron scattering experiments \cite{Sidis}.
Namely, in contrast to the naive expectation, paramagnons do not
always play a central role in the occurrence of triplet
superconductivity in the solid state.
It is still a puzzling and challenging problem to clarify
a key issue to determine the Cooper-pair symmetry.

One possible scenario is based on the Hund's rule coupling,
which stabilizes the local triplet pair composed of a couple of
electrons between different orbitals.
Then, the triplet superconductivity is naively expected to occur,
but in such a local picture, it is questionable whether
the anisotropic Cooper-pair is stabilized or not.
This point has casted a serious doubt on the scenario,
but in the course of the investigation on UPt$_3$,
it has been pointed out that an odd-parity triplet pair
is realized only when the inversion center exists
external to the $f$-shell ions \cite{Anderson}.
This has been further examined by the detailed
group-theoretical analysis \cite{Appel} and
the estimation of pairing potential with the use
of band-calculation results \cite{Norman}.
However, the discussions have been in a phenomenological level and
it is highly required that the triplet pair induced by the Hund's rule
coupling should be investigated from the microscopic viewpoint.

In this Letter, we attempt to gain an insight into triplet pairing
induced by Hund's rule coupling.
First we reconsider the symmetry argument on triplet pair in the
weak-coupling limit.
It is again found that the odd-parity triplet pair induced by
the Hund's rule interaction occurs only for the non Bravais lattice.
In order to confirm this result from the microscopic viewpoint,
we carefully analyze an orbital degenerate model on the two-dimensional
(2D) square and honeycomb lattices.
It is emphasized that the pair state is determined unambiguously
by diagonalizing the pair correlation function
based on the concept of off-diagonal long-range order \cite{Yang}.
Then, we can visualize the pair wavefunction in an unbiased manner,
clearly indicating that the odd-parity triplet pair in the FM phase
occurs as an out-of-phase combination of local triplets
induced by Hund's rule coupling in the 2D honeycomb lattice
with two sites in the unit cell.


Let us start our discussion on the Cooper-pair amplitude
in the lattice with inversion symmetry.
The pair operator composed of a couple of electrons is defined
in the second quantized form as
\begin{equation}
  \label{Eq:cooperpair}
  \phi_{{\bf i}\mu\alpha\sigma,\,{\bf j}\nu\beta\tau}
  =c_{{\bf i}\mu\alpha\sigma} c_{{\bf j}\nu\beta\tau},
\end{equation}
where $c_{{\bf i}\mu\alpha\sigma}$ is an annihilation operator
of an electron with spin $\sigma$ in the orbital $\alpha$
on the site $\mu$ included in the unit cell ${\bf i}$.
We consider the Fourier transform of $c_{{\bf i}\mu\alpha\sigma}$ as
\begin{equation}
  \label{Eq:fourier}
  c_{{\bf i}\mu\alpha\sigma}
  =\sum_{n,\bm{k}}
  e^{i{\bm k}\cdot{\bm R}_{\bf i}}
  \Lambda_{\mu\alpha}^{\bm{k}n}a_{{\bm k}n\sigma},
\end{equation}
where $a_{\bm{k}n\sigma}$ is an annihilation operator for
an electron with momentum $\bm{k}$ and spin $\sigma$
in the $n$-th band.
Note that $\Lambda_{\mu\alpha}^{\bm{k}n}$ is the matrix element
of the unitary transformation between orbitals and bands.

Now let us consider the Cooper pair of which the total momentum
is equal to zero in the weak-coupling limit.
Then, the Cooper-pair amplitude is given by
\begin{equation}
 \label{Eq:pair}
 \! \langle \phi_{{\bf i}\mu\alpha\sigma,\,{\bf j}\nu\beta\tau}
 \rangle \!=\! \sum_{n,\bm{k}}
 e^{i{\bm k}\cdot({\bm R}_{\bf i}-{\bm R}_{\bf j})}
 \Lambda^{\bm{k}n}_{\mu\alpha}\Lambda^{-\bm{k}n}_{\nu\beta}
 \langle a_{\bm{k} n \sigma} a_{\bm{-k} n \tau} \rangle.
\end{equation}
Note that the Cooper pair is formed only by a couple of electrons
on the same electronic band, since the pair composed of electrons
in different bands is not stable in general in the weak-coupling limit.

\begin{figure}[t]
\includegraphics[width=1.0\linewidth]{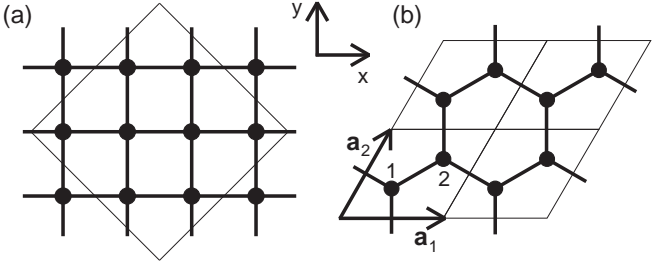}
\caption{(a) Two-dimensional square and (b) honeycomb lattices.
Thin lines in (a) denotes 8-site cluster, while in (b),
they indicate 8-site cluster composed of 4 unit cells
including two sites, 1 and 2.
Thick arrows define vectors ${\bf a}_1$ and ${\bf a}_2$.}
\end{figure}

Here we introduce an inversion operator ${\cal P}$, which acts
generally on the electron operator as
\begin{equation}
 \label{Eq:inversion}
 {\cal P} c_{{\bf i}\mu\alpha\sigma} {\cal P}^{-1}
  =P_{\alpha}c_{{\bf i}\nu\alpha\sigma},
\end{equation}
where $P_{\alpha}$ denotes the parity for the inversion
of atomic orbital, depending on the angular momentum.
In this paper, we assume the pairing in the orbitals with
the same angular momentum, indicating $P_{\alpha}$=$P_{\beta}$.

For the Bravais lattice in which one site is included in
the unit cell, typically the square lattice shown in Fig.~1(a),
the inversion center is located on the site.
Thus, we can simply suppress the indices $\mu$ and $\nu$
in this case.
Combing Eqs.~(\ref{Eq:fourier}) and (\ref{Eq:inversion}),
we obtain
\begin{equation}
 \label{Eq:invb}
 \Lambda_{\alpha}^{\bm{k}n}
 =P_{\alpha} \Lambda_{\alpha}^{\bm{-k}n},
\end{equation}
for the Bravais lattice.
On the other hand, for the non Bravais lattice,
$\mu$ is not equal to $\nu$ in Eq.~(\ref{Eq:inversion}).
As a typical example, here we consider
the honeycomb lattice composed of unit cells
including two sites, as shown in Fig.~1(b).
In this case, the inversion center is located at the center
of two sites and we obtain
\begin{equation}
 \label{Eq:invnb}
 \Lambda_{1\alpha}^{\bm{k}n}
 =P_{\alpha} \Lambda_{2\alpha}^{\bm{-k}n}.
\end{equation}

First we consider the Cooper pair in the Bravais lattice.
Note again that the index $\mu$ is suppressed in this case.
After algebraic calculations, spin singlet (S) and triplet
(T) pairs are, respectively, given by
\begin{eqnarray}
 \langle \phi^{\rm S}_{{\bf i}\alpha,\,{\bf j}\beta} \rangle
 =\sum_{n,\bm{k}}
 \cos [\bm{k} \cdot (\bm{R}_{\bf i}-\bm{R}_{\bf j})]
 \Lambda_{\alpha}^{\bm{-k}n} \Lambda_{\beta}^{\bm{k}n}
 S_{\bm{k}n},
\end{eqnarray}
and
\begin{equation}
 \langle \phi^{{\rm T},z}_{{\bf i}\alpha,\,{\bf j}\beta} \rangle
 =\sum_{n,\bm{k}}
 i \sin [\bm{k} \cdot (\bm{R}_{\bf i}-\bm{R}_{\bf j})]
 \Lambda_{\alpha}^{\bm{-k}n} \Lambda_{\beta}^{\bm{k}n}
 T^{z}_{\bm{k}n},
\end{equation}
where
$S_{\bm{k}n}
=(\langle a_{\bm{k} n \uparrow} a_{\bm{-k} n \downarrow} \rangle
-\langle a_{\bm{k} n \downarrow} a_{\bm{-k} n \uparrow} \rangle)
/\sqrt{2}$,
$T^{+1}_{\bm{k}n}
=\langle a_{\bm{k} n \uparrow} a_{\bm{-k} n \uparrow} \rangle$,
$T^{0}_{\bm{k}n}=
(\langle a_{\bm{k} n \uparrow} a_{\bm{-k} n \downarrow} \rangle
+\langle a_{\bm{k} n \downarrow} a_{\bm{-k} n \uparrow} \rangle)
/\sqrt{2}$,
and
$T^{-1}_{\bm{k}n}=
\langle a_{\bm{k} n \downarrow} a_{\bm{-k} n \downarrow} \rangle$.
As expected, spin singlet and triplet pairs have
even- and odd-parity, respectively.

Note here that the magnitude of on-site pair-amplitude vanishes
for the odd-parity triplet pair, while for the singlet case,
it depends on the symmetry of Cooper pair.
As is well known, the Hund's rule coupling stabilizes
the local triplet pair composed of electrons in different orbitals,
inevitably leading to the orbital antisymmetric pair with
finite on-site amplitude \cite{Hotta}.
However, such an even-parity triplet pair does not appear
in the weak-coupling limit.
In order to make this point clear, it is instructive to
consider the on-site pair-amplitude in the two-band model
on the Bravais lattice.
After simple algebraic calculations, we obtain
$\langle c_{{\bf i}\alpha\sigma} c_{{\bf i}\beta\sigma} \rangle$
=$\sum_{\bm{k}}\langle a_{\bm{k} 1 \sigma} a_{\bm{-k} 2 \sigma} \rangle$
for $\alpha$$\ne$$\beta$.
Namely, the triplet pair induced by the Hund's rule coupling
should be composed of electrons on different bands,
which is not stable in the weak-coupling limit \cite{comment1}.
We conclude that for the Bravais lattice, odd-parity triplet pair
cannot be induced by the Hund's rule coupling.

\begin{figure}[t]
\includegraphics[width=0.8\linewidth]{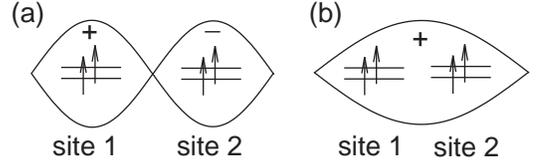}
\caption{Schematic views for wavefunctions of
(a) odd-parity and (b) even-parity triplet pairs
in the honeycomb lattice.}
\end{figure}

Next we turn our attention to the pair formation
in the non Bravais lattice, typically the honeycomb lattice.
For simplicity, we consider only the on-site pair amplitude.
Using Eqs.~(\ref{Eq:pair}) and (\ref{Eq:invnb}),
for even-parity singlet and odd-parity triplet pairs, we obtain
\begin{equation}
 \langle \phi^{\rm S}_{{\bf i}1\alpha,\,{\bf i}1\beta} \rangle
 \!=\!(1/2)\sum_{n,\bm{k}}(
 \Lambda_{1\alpha}^{\bm{k}n}\Lambda_{1\beta}^{\bm{-k}n}
 +\Lambda_{2\alpha}^{\bm{-k}n}\Lambda_{2\beta}^{\bm{k}n})
 S_{\bm{k}n},
\end{equation}
and
\begin{equation}
 \langle \phi^{{\rm T},z}_{{\bf i}1\alpha,\,{\bf i}1\beta}
 \rangle
 \!=\!(1/2)\sum_{n,\bm{k}}(
 \Lambda_{1\alpha}^{\bm{k}n}\Lambda_{1\beta}^{\bm{-k}n}
 -\Lambda_{2\alpha}^{\bm{-k}n}\Lambda_{2\beta}^{\bm{k}n})
 T^{z}_{\bm{k}n},
\end{equation}
respectively.
Note that in general,
$\langle \phi^{{\rm T},z}_{{\bf i}1\alpha,\,{\bf i}1\beta}
\rangle$ does $not$ vanish.
In order to understand this point, it is useful to remark the relation
\begin{equation}
 \label{Eq:odd}
 \langle \phi^{{\rm T},z}_{{\bf i}1\alpha,\,{\bf i}1\beta}
 \rangle
 =-\langle \phi^{{\rm T},z}_{{\bf i}2\alpha,\,{\bf i}2\beta}
 \rangle.
\end{equation}
The pair satisfying this relation is composed of the local triplet
pairs in the unit cell in an out-of-phase manner,
as schematically shown in Fig.~2(a).
Then, the odd-parity triplet pair induced by the
Hund's rule coupling can appear in the non Bravais lattice.

We believe that the above symmetry argument is useful, but
it mentions nothing about the stability of the odd-parity pair.
For instance, the in-phase combination of local triplet pairs
with even-parity [see Fig.~2(b)] seems to dominate the
odd-parity pair at a first glance, but the above discussion
does not conclude which pair is stabilized.
In order to complete the discussion, we cannot avoid to
carry out some explicit calculations in an appropriate model.
However, if we apply mean-field approximations on
the model in which attractive interactions are
introduced just by hand, we may lose essential points.
Thus, in this paper, we adopt the orbital degenerate Hubbard model,
which is widely believed to be a standard model
for strongly correlated electron systems.
Furthermore, in order to obtain unbiased results, we resort to
the numerical method such as exact diagonalization.
For the purpose to analyze the paring symmetry in the numerical
calculation, we apply the method of the optimization of
pair correlation function \cite{Hotta}.

Here we take the $e_{\rm g}$-orbital
Hubbard model as \cite{comment2}
\begin{eqnarray}
 H &=&
 -\sum_{\langle {\bf i}\mu,{\bf j}\nu \rangle}\sum_{\sigma,\alpha,\beta}
 t^{\bf v}_{\alpha\beta}
 (c^{\dag}_{{\bf i}\mu\alpha\sigma}c_{{\bf j}\nu\beta\sigma}+{\rm h.c.})
 \nonumber \\
 &+& U \sum_{{\bf i}, \mu, \alpha}
 \rho_{{\bf i}\mu\alpha\uparrow} \rho_{{\bf i}\mu\alpha\downarrow}
 +U'/2 \sum_{{\bf i}, \mu} \sum_{\alpha \ne \beta}
 \rho_{{\bf i}\mu\alpha} \rho_{{\bf i}\mu\beta}
 \nonumber \\
 &+& J/2 \sum_{{\bf i}, \mu} \sum_{\sigma,\sigma',\alpha \ne \beta} 
 c_{{\bf i}\mu\alpha\sigma}^{\dag} c_{{\bf i}\mu\beta\sigma'}^{\dag}
 c_{{\bf i}\mu\alpha\sigma'} c_{{\bf i}\mu\beta\sigma} \nonumber \\
 &+& J' \sum_{{\bf i},\mu} \sum_{\alpha \ne \beta} 
 c_{{\bf i}\mu\alpha\uparrow}^{\dag} c_{{\bf i}\mu\alpha\downarrow}^{\dag}
 c_{{\bf i}\mu\beta\downarrow} c_{{\bf i}\mu\beta\uparrow},
\end{eqnarray}
where $\langle {\bf i}\mu,{\bf j}\nu \rangle$ denotes a pair of
nearest-neighbor sites,
${\bf v}$ indicates a vector connecting the nearest-neighbor sites,
$\alpha$ ($\beta$) denotes $x^2$$-$$y^2$ ($3z^2$$-$$r^2$) orbital,
$\rho_{{\bf i}\mu\alpha\sigma}$=
$c_{{\bf i}\mu\alpha\sigma}^{\dag} c_{{\bf i}\mu\alpha\sigma}$, and
$\rho_{{\bf i}\mu\alpha}$=$\sum_{\sigma}\rho_{{\bf i}\mu\alpha\sigma}$.
In the first term, $t^{\bf v}_{\alpha\beta}$ is a hopping amplitude
depending on orbitals, hopping directions, and lattice type.
Hereafter $t^{\bf y}_{\alpha\alpha}$ along the ${\bf y}$ direction
(see Fig.~1) is taken as the energy unit.
In the Coulomb interaction terms, $U$, $U'$, $J$, and $J'$ denote
intra-orbital, inter-orbital, Hund's rule, and pair-hopping
interactions, respectively.
Note the relations of $U$=$U'$+$J$+$J'$ and $J$=$J'$.

\begin{figure}[t]
\includegraphics[width=1.0\linewidth]{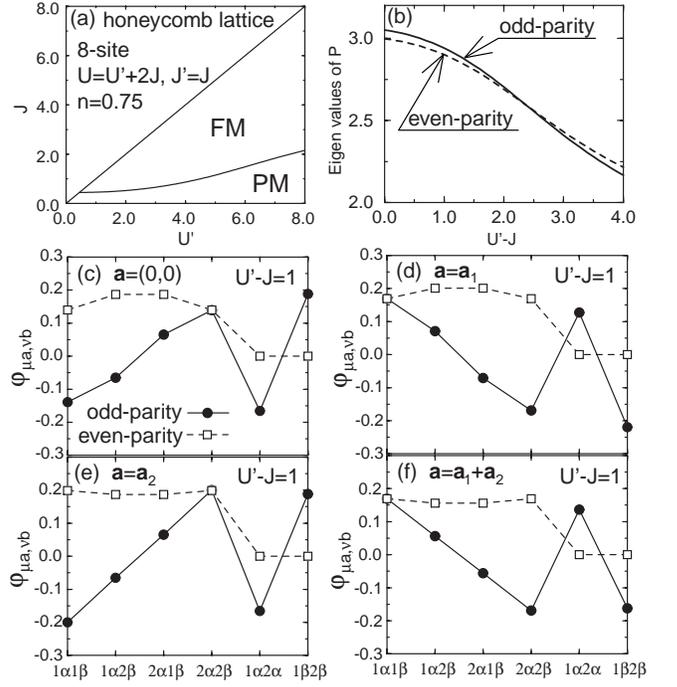}
\caption{(a) Ground-state phase diagram for 8-site honeycomb lattice
with $n$=0.75. Note that the region for $J$$>$$U'$ is unphysical.
(b) Maximum and second maximum eigen-values of $P$ as a function
of $U'$$-$$J$ in the FM phase.
Coefficients of the odd- and even-parity triplet pair correlation
functions for (c) ${\bf a}$=(0,0), (d) ${\bf a}$=${\bf a}_1$,
(e) ${\bf a}$=${\bf a}_2$, and (f) ${\bf a}$=${\bf a}_1$+${\bf a}_2$,
for $U'$$-$$J$=1.}
\end{figure}

First let us briefly discuss the ground-state phase diagrams
for two types of 2D lattices with 8 sites (see Fig.~1).
In Fig.~3(a), the phase diagram for the 8-site honeycomb lattice is
shown for $n$=0.75.
When $J$ is increased, the FM phase appears,
since it can gain the kinetic energy for large $J$.
This is quite general in the orbital degenerate model.
For $n$=1 (not shown here), the FM phase is also found,
but the region becomes narrow:
It does not touch the line of $J$=$U'$.
For the 2D square lattice with 8 sites, the phase diagram for $n$=1.5
is quite similar to Fig.~3(a) (see Ref.~\cite{Hotta}).
For $n$=1, the phase diagram is also similar to that of
the honeycomb lattice with $n$=1.
For $n$=0.75, the FM phase disappears in the square lattice.

Now we focus on the pair in the FM phase to clarify
the symmetry of triplet pair induced by the Hund's rule coupling.
For the purpose, we measure the triplet pair correlation function,
defined as $P$=$\langle \Phi \Phi^{\dag} \rangle$, where
$\Phi$ is a triplet pair operator given by a linear combination
of $\phi_{{\bf i}\mu\alpha, \, {\bf j}\nu\beta}^{{\rm T},z}$.
The coefficient of each component includes information about
pairing symmetry.
The pair wavefunction should be determined
as the eigen function with the maximum eigenvalue of $P$ \cite{Hotta}.
The pairing state thus determined,
called ``optimized pair'', is the most
probable candidate for the Cooper pair in actual systems.

Before proceeding to the exhibition of numerical results,
let us briefly discuss the meaning of the optimized pair
from the conceptual viewpoint of
off-diagonal long-range order \cite{Yang}.
In principle, the occurrence of superconductivity is detected
when the maximum eigenvalue $\rho_{\rm max}$ of $P$ becomes
the order of $N$, where $N$ is the number of electrons.
In a small cluster calculation, the possible superconducting pair
state should be defined by the eigenstate with $\rho_{\rm max}$,
but in order to prove the existence of off-diagonal long-range order,
it is necessary to show that $\rho_{\rm max}$ actually
becomes the order of $N$ with increasing the system size.
However, at some cluster size, a possible pairing state is
determined without ambiguity by the eigenstate with $\rho_{\rm max}$.

In Fig.~3(b), we show the eigenvalues of $P$ in the FM phase
of the honeycomb lattice for $n$=0.75.
For small $U'$, the eigenstate of $P$ corresponding to
$\rho_{\rm max}$ exhibits odd-parity,
while the even-parity state appears for large value of $U'$
\cite{Hotta2}.
As discussed below, the symmetry of those pairing states are
consistent with the schematic views shown in Figs.~2(a) and (b).
In the fully spin-polarized phase, the relevant interaction
is $U'$$-$$J$.
Then, small $U'$$-$$J$ corresponds to the ``weak-coupling'' region
in the FM phase.
In that sense, it seems natural that the odd-parity state
appears for small $U'$$-$$J$, while the even-parity occurs
for large $U'$$-$$J$.

In order to visualize the paring symmetry, we explicitly
express the pair operators as
\begin{eqnarray}
 \label{Eq:pairfunction}
 \Phi &=& \sum_{\bf i,a}
 (\varphi^{\bf a}_{1\alpha,1\beta}
 c_{{\bf i}1 \alpha \uparrow} c_{{\bf i+a}1 \beta \uparrow}
 +\varphi^{\bf a}_{1\alpha,2\beta}
 c_{{\bf i}1 \alpha \uparrow} c_{{\bf i+a}2 \beta \uparrow} \nonumber \\
 &+& \varphi^{\bf a}_{2\alpha,1\beta}
 c_{{\bf i}2 \alpha \uparrow} c_{{\bf i+a}1 \beta \uparrow}
 +\varphi^{\bf a}_{2\alpha,2\beta}
 c_{{\bf i}2 \alpha \uparrow} c_{{\bf i+a}2 \beta \uparrow} \nonumber \\
 &+& \varphi^{\bf a}_{1\alpha,2\alpha}
 c_{{\bf i}1 \alpha \uparrow} c_{{\bf i+a}2 \alpha \uparrow}
 +\varphi^{\bf a}_{1\beta,2\beta}
 c_{{\bf i}1 \beta \uparrow} c_{{\bf i+a}2 \beta \uparrow}),
\end{eqnarray}
where ${\bf a}$ is a vector connecting two unit cells.
See Fig.~1(b) for the definitions of ${\bf a}$
in the honeycomb lattice.

In Figs.~3(c)-(f), we depict the coefficients
$\varphi^{\bf a}_{\mu{\rm a},\nu{\rm b}}$ for each ${\bf a}$
and $U'$$-$$J$=1 \cite{note}.
The symmetry argument in the weak-coupling limit has suggested that
the odd-parity triplet pair should satisfy the relations,
$\varphi^{\bf a}_{1\alpha,1\beta}$=$-\varphi^{\bf a}_{2\alpha,2\beta}$
and
$\varphi^{\bf a}_{1\alpha,2\beta}$=$-\varphi^{\bf a}_{2\alpha,1\beta}$
\cite{comment3}.
In Fig.~3(c), these relations are actually satisfied
in the pair correlation inside the unit cell for ${\bf a}$=$(0, 0)$,
indicating $unbiased$ evidence for the odd-parity triplet pair.
As shown in Figs.~3(d)-(f), the above relations for ${\bf a}$=(0,0)
are also observed for ${\bf a}$$\ne$(0,0) and the pair amplitudes
for ${\bf a}$$\ne$(0,0) are in the same order as those
for ${\bf a}$=(0,0).
These results indicate that the triplet pair spatially extends
to gain the kinetic energy by keeping the odd-parity symmetry.
Here note an in-phase relation in the coefficients
for ${\bf a}_2$ direction, while an out-of-phase relation
for ${\bf a}_1$ direction, suggesting a node structure
between the unit cells along ${\bf a}_1$ direction.
Namely, the $p$-wave-like pair is appearing in this case.
Note also another relation
$\varphi^{\bf a}_{1\alpha,2\alpha} \approx
-\varphi^{\bf a}_{1\beta,2\beta}$ for the odd-parity pair,
indicating the node of the pair wavefunction in the orbital space.

Finally, we briefly comment on actual compounds from the
present result.
First, the odd-parity triplet pair in the FM phase induced
by the Hund's rule coupling is expected to occur in some $f$-electron
compounds such as UGe$_2$ and URhGe, although
further quantitative calculations are needed based on more
realistic models.
Second, the triplet pair in ruthenate may not be induced
by the local Hund's rule interaction, since the ruthenate
is described by the $t_{\rm 2g}$-orbital degenerate Hubbard
model on the 2D square lattice.
Rather we may consider a scenario based on the pairing induced by
the effective interaction among quasi-particles in the Fermi-liquid
theory \cite{Nomura}.

In summary, we have discussed microscopic aspects of 
odd-parity triplet pair.
The weak-coupling analysis has suggested that
the odd-parity triplet pair induced by the Hund's rule coupling
does not appear in the Bravais lattice.
We have carefully analyzed the triplet pair wavefunction
in the FM phase on the 2D square and honeycomb lattices.
It has been clarified that the odd-parity triplet pair induced
by the local Hund's coupling occurs only for
the non Bravais lattice.

The authors are separately supported by the Grant-in-Aid for
Scientific Research from Japan Society for the Promotion of Science.



\begin{thebibliography}{99}

\bibitem{UPt3}
G. R. Stewart {\it et al.}, Phys. Rev. Lett. {\bf 52}, 679 (1984).

\bibitem{Tou}
H. Tou {\it et al.}, Phys. Rev. Lett. {\bf 77}, 1374 (1996).

\bibitem{Saxena}
S. S. Saxena {\it et al.}, Nature (London) {\bf 406}, 587 (2000).

\bibitem{Aoki}
D. Aoki {\it et al.}, Nature (London) {\bf 413}, 613 (2001).

\bibitem{Ishida}
K. Ishida {\it et al.}, Phys. Rev. Lett. {\bf 89}, 037002 (2002).

\bibitem{Maeno}
Y. Maeno {\it et al.}, Nature (London) {\bf 372}, 532 (1994).

\bibitem{Ishida2}
K. Ishida {\it et al.}, Nature (London) {\bf 396}, 658 (1998).

\bibitem{ZrZn2}
C. Pfleiderer {\it et al.}, Nature (London) {\bf 412}, 58 (2001).

\bibitem{Sidis}
Y. Sidis {\it et al.}, Phys. Rev. Lett. {\bf 83}, 3320 (1999).

\bibitem{Anderson}
P. W. Anderson, Phys. Rev. B {\bf 32}, 499 (1985).

\bibitem{Appel}
J. Appel and P. Hertel, Phys. Rev. B {\bf 35}, 155 (1987).

\bibitem{Norman}
M. R. Norman, Phys. Rev. Lett. {\bf 72}, 2077 (1994);
Phys. Rev. B {\bf 50}, 6904 (1994).

\bibitem{Yang}
C. N. Yang, Rev. Mod. Phys. {\bf 34}, 694 (1962).

\bibitem{Hotta}
T. Hotta and K. Ueda, Phys. Rev. B {\bf 67}, 104518 (2003).

\bibitem{comment1}
In the strong-coupling case, pair formation occurs
in the whole Fermi sphere. Namely, the pair can be formed by
electrons in different bands, leading to hybridization of
non-degenerate bands in the pair wavefunction.

\bibitem{comment2}
When we apply this model to $f$-electron systems, we need to
pay attention to the meanings of ``spin'' and ``orbital'' of $f$
electron, as discussed in Ref.~\cite{Hotta}.
Note also that it is allowed to define an $e_{\rm g}$-orbital
Hubbard model on the honeycomb lattice, since local orbitals should
be determined from the effects of ligand ions, not directly related
to the whole symmetry of the lattice.
Unfortunately, it is difficult to show actual materials described by
the present Hamiltonian, but this is the minimal model
with degenerate orbitals in the non Bravais lattice.

\bibitem{Hotta2}
T. Hotta and K. Ueda, to appear in J. Magn. Magn. Matter.

\bibitem{note}
The qualitative feature of the triplet pair in the FM phase
does not depend on the specific value of $U'$$-$$J$.

\bibitem{comment3}
On the other hand, the even-parity triplet pair satisfies
relations $\varphi^{\bf a}_{1\alpha,1\beta}$=
$\varphi^{\bf a}_{2\alpha,2\beta}$ and
$\varphi^{\bf a}_{1\alpha,2\beta}$=
$\varphi^{\bf a}_{2\alpha,1\beta}$=0.

\bibitem{Nomura}
T. Nomura and K. Yamada, J. Phys. Soc. Jpn. {\bf 71},
1993 (2002).

\end{thebibliography}
\end{document}